\documentclass[%
 reprint,
 amsmath,amssymb,
 aps,
]{revtex4-2}

\usepackage{graphicx}
\usepackage{dcolumn}
\usepackage{bm}

\usepackage{color}

\definecolor{ccomments}{rgb}{1,0.2,0.2}

\definecolor{cmissing}{rgb}{1,0.5,0.3}

\begin{document}


\title{Spontaneous filament formation and network self-assembly via active phase
separation}

\author{Elena Lucas$^1$, Varun Venkatesh$^1$, Amin Doostmohammadi$^{1 \dagger}$}
\address{ $^1$ Niels Bohr Institute, University of Copenhagen, Denmark\\
$\dagger$: doostmohammadi@nbi.ku.dk.}

\begin{abstract}
 We introduce Active Model H$^{-}$, a scalar phase-field model of non-equilibrium phase separation in the overdamped hydrodynamic limit, and show that it produces filamentous networks with enhanced connections and arrested coarsening, a morphology not observed in existing scalar active field theories. Isolated filaments are unstable to spontaneous bending, and the resulting curvature drives self-propulsion towards the convex side, causing them to collide and assemble into a percolating network. The networks are locally non-equilibrium, with anomalously curved edges, disordered vertex angles and heterogeneous topological charge, yet recover Lewis's law macroscopically. Under noise they show balanced break and reform dynamics.
\end{abstract}


\maketitle

\section*{Introduction}

Networks of filamentous domains are a recurring morphology in out-of-equilibrium soft matter
systems. 
Wormlike micelles form dense, entangled networks that coexist with a dilute phase and continuously break and recombine~\cite{wormlike_micelles_review, CatesCandau1990, Khatory1993}. In polymer and protein solutions with viscoelastic asymmetry between the components, viscoelastic phase separation produces transient network structures of the slower, more elastic component before coarsening proceeds~\cite{Tanaka1993, Tanaka2000, TanakaNishikawa2005, Tanaka2009, Tanaka2022}. In emulsions quenched far from equilibrium, interfacially jammed colloids arrest a bicontinuous morphology indefinitely~\cite{Stratford2005, Herzig2007}. In each case the network is not a thermodynamic minimum but a state selected by the competition between phase separation and some arresting mechanism. Understanding which minimal ingredients are sufficient to produce and sustain such states is an open problem in non-equilibrium physics.

Liquid--liquid phase separation provides a natural starting point. The Cahn--Hilliard equation and its active extensions have proven to be powerful frameworks for describing morphological transitions in binary mixtures driven out of equilibrium~\cite{CahnHilliard_freeenergy1958, Bray2002, fluidmixtures_review_CatesTjhung2018, CatesTailleur2015, AMB+_review_Cates2025, AMB-_ToffenettiFrey2026}. Active Model B$+$ (AMB$+$), the leading scalar theory, introduces non-equilibrium terms~\cite{AMB_original_Wittkowski2014,  AMB+_original_Tjhung2018, Nardini2017} that arrest Ostwald ripening and produce microphase separation and foam-like structures~\cite{AMB+_original_Tjhung2018, AMB+_review_Cates2025, SinghCates2019}. However, the AMB$+$ morphology is dominated by wide, blobby junctions, and connections appear only as weak bridges of low density. Genuinely connected filamentous networks, in which every junction is composed of bulk dense phase, have recently been obtained by coupling a scalar concentration field to a tensorial nematic order parameter~\cite{activefluids_formnetworks}, building on earlier scalar--tensor treatments of phase-separating active liquid crystals~\cite{Blow2014} and on experiments on active liquid interfaces~\cite{Adkins2022}. This, however, comes at the cost of significant additional complexity, and it requires an explicit alignment field that has no scalar counterpart. Whether a purely scalar theory can produce truly connected networks has remained an open question.

Here we show that it can. We introduce Active Model H$^{-}$, derived as the overdamped limit of Active Model H~\cite{AMH_original_Tiribocchi2015,  SinghCates2019, Caballero2025} with a curvature-motivated interfacial stress~\cite{Anderson1998}, and demonstrate that it produces fully connected filamentous networks with arrested coarsening across a broad region of parameter space. The model requires neither an alignment field nor any additional order parameter. The arrested state arises from a single emergent mechanism: filaments are unstable to spontaneous bending, and the resulting curvature drives self-propulsion towards the convex side of the interface, causing isolated filaments to collide and assemble into a percolating network. These arrested networks have a non-equilibrium morphology, with anomalously curved edges, persistent vertex-angle disorder, and spatially heterogeneous topological charge, yet Lewis's law is recovered at the macroscopic scale~\cite{Lewis1928, WeaireRivier1984}. Finally, under additive noise the network breaks and reforms continuously, maintained by the same mechanism that assembles it.

\section*{Model}
\begin{figure*}[!t]
\begin{center}
\includegraphics[width=165mm]{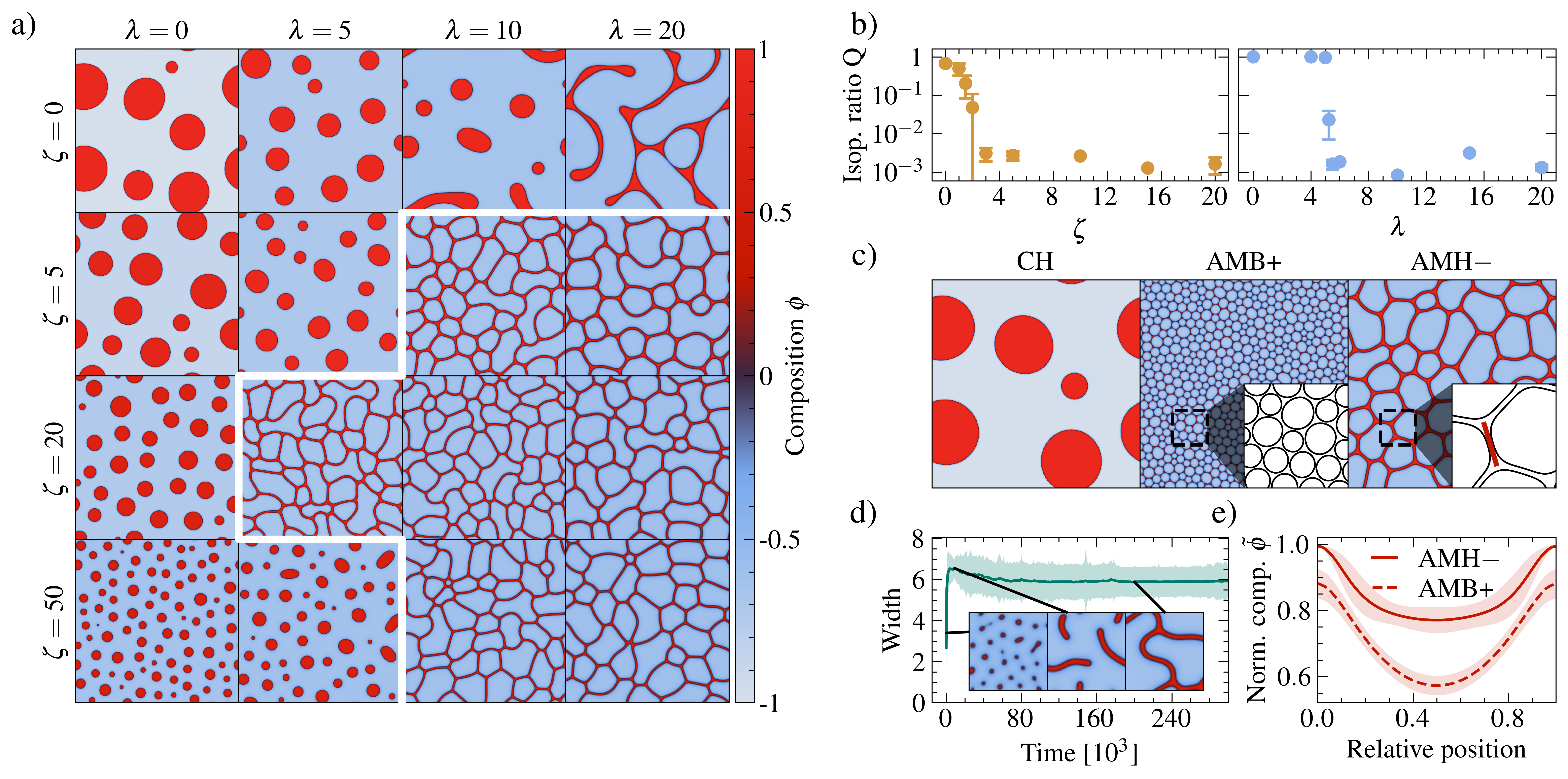}
    \caption{Active Model H$^{-}$ produces connected filamentous networks with arrested coarsening.
    (a) Phase diagram at $t=500\times10^3$ in the $(\lambda, \zeta)$ parameter space; the white line delineates the region of percolating networks.
    (b) Network isoperimetric ratio as a function of $\zeta$ (left, with $\lambda=10$) and $\lambda$ (right, $\zeta=50$) separately, taken from three samples at $t=500\times10^3$.
    (c) Snapshots at $t=1000\times10^3$ of the passive Cahn--Hilliard equation, AMB$+$, and Active Model H$^{-}$ at the representative parameters $(\lambda, \zeta) = (10, 5)$. The insets show a zoomed-in view of the contours at $\phi_\mathrm{mid} = (\phi_+ + \phi_-)/2$. (d) Mean filament width with standard deviation as a function of time for a representative parameter set $(\lambda, \zeta) = (10, 50)$, with snapshots (inset) of filaments at $t=(0.3, 9, 200)\times10^3$.
    (e) The normalised average field $\tilde{\phi} = (\phi-\phi_-)/(\phi_+-\phi_-)$ with standard deviation along the filament, normalised to its length, as marked by the red line in the contour inset of (c). Data were averaged at $t=(500,750,1000)\times10^3$.}
    \label{fig:phase_diagram}
    \label{fig.1}
\end{center}
\end{figure*}

We consider a binary mixture described by a conserved scalar composition field $\phi(\mathbf{r},
t)$. The free energy functional takes the standard Landau--Ginzburg form
\begin{equation}
    \mathcal{F}[\phi] = \int \left[ f(\phi) + \frac{k}{2}(\nabla\phi)^2
    \right] d\mathbf{r},
    \label{eq:freeenergy}
\end{equation}
with a symmetric double-well bulk free energy $f(\phi) = \gamma(\phi^4/4 - \phi^2/2)$
and an interfacial penalty controlled by $k$. The exchange chemical potential is
$\mu = f'(\phi) - k\Delta\phi$.

To model active driving, we follow the Active Model H framework~\cite{AMH_original_Tiribocchi2015}
in which activity enters through a deviatoric interfacial stress
\begin{equation}
    \Sigma_{ij} = -\left(\partial_i\phi\,\partial_j\phi -
    \frac{1}{2}|\nabla\phi|^2\delta_{ij}\right).
    \label{eq:stress}
\end{equation}
This encodes the local curvature of the phase boundary through the traceless
combination of field gradients.

Taking the overdamped limit appropriate
for a high-viscosity medium, the velocity
field reduces to $\mathbf{v} \propto \nabla\cdot\Sigma$. Using Eq.~(\ref{eq:stress}), the divergence of the stress is $\nabla\cdot\Sigma = -\Delta\phi\,\nabla\phi$, so that the advective contribution $\phi\mathbf{v}$ to the flux is proportional to $-\phi\,\Delta\phi\,\nabla\phi$. Substituting this into the
Cahn--Hilliard equation, with the chemical
potential $\mu_\lambda$ of Ref.~\cite{AMB_original_Wittkowski2014}, yields the
governing equations
\begin{eqnarray}
    \partial_t\phi &=& -\nabla\cdot\boldsymbol{J}, \label{eq:continuity}\\
    \boldsymbol{J} &=& -M\nabla\mu_\lambda + \zeta\left(\phi\,\Delta\phi\,\nabla\phi\right), \label{eq:AMH}\\
    \mu_\lambda &=& f'(\phi) - k\Delta\phi + \lambda|\nabla\phi|^2.
    \label{eq:mu}
\end{eqnarray}
We refer to this as Active Model H$^{-}$. Here $M$ is the mobility and the activity coefficient $\zeta$ absorbs the positive friction prefactor of the overdamped limit, so that for $\zeta>0$ the advective flux is directed from the concave towards the convex side of a curved interface. The model has two activity parameters: $\lambda$, which modifies the effective interfacial tension~\cite{AMB_original_Wittkowski2014} and drives domain elongation; and $\zeta$, the overdamped advection strength, which generates circulating currents and arrests coarsening. Equation~(\ref{eq:AMH}) can be written as AMB$+$ with a $\phi$-dependent coefficient $\zeta \to \zeta\phi$, which increases
the nonlinear order by one. We begin with the deterministic model described above and later incorporate noise by adding a stochastic flux $\boldsymbol{J}_\sigma=\sigma \,\boldsymbol{\Lambda}$ to $\boldsymbol{J}$, with $\boldsymbol{\Lambda}$ denoting a Gaussian white noise vector and $\sigma$ the noise strength.

Simulations are performed on a periodic two-dimensional square lattice of side length $512$, except for the isolated-filament study (side length $256$) and the network morphology analysis (side length $2048$). Time integration is carried out with an explicit Euler scheme using nine-point finite-difference stencils~\cite{stencils_Kumar2004}. The parameters are set to $\gamma = 1$, $k = 2$ and $M = 1$, while $\lambda$ is varied in $[0,20]$ and $\zeta$ in $[0,50]$.
Spatial and temporal resolutions are $\Delta x = 1$ and $\Delta t = 0.01$, refined to
$\Delta x = 0.5$ and $\Delta t = 0.002$ at high activity ($\zeta\geq5$ and $\lambda\geq10$). The average concentration is set to $\phi_0=-0.4$ unless specified otherwise. We define the dense phase as $\phi>(\phi_+ +\phi_-)/2 $ where $\phi_+$  and $\phi_-$ are the maximum and minimum of the phase field in the long time limit.

\subsection*{A new morphological state: connected filamentous networks}

\begin{figure*}
    \begin{center}
    \includegraphics[width=165mm]{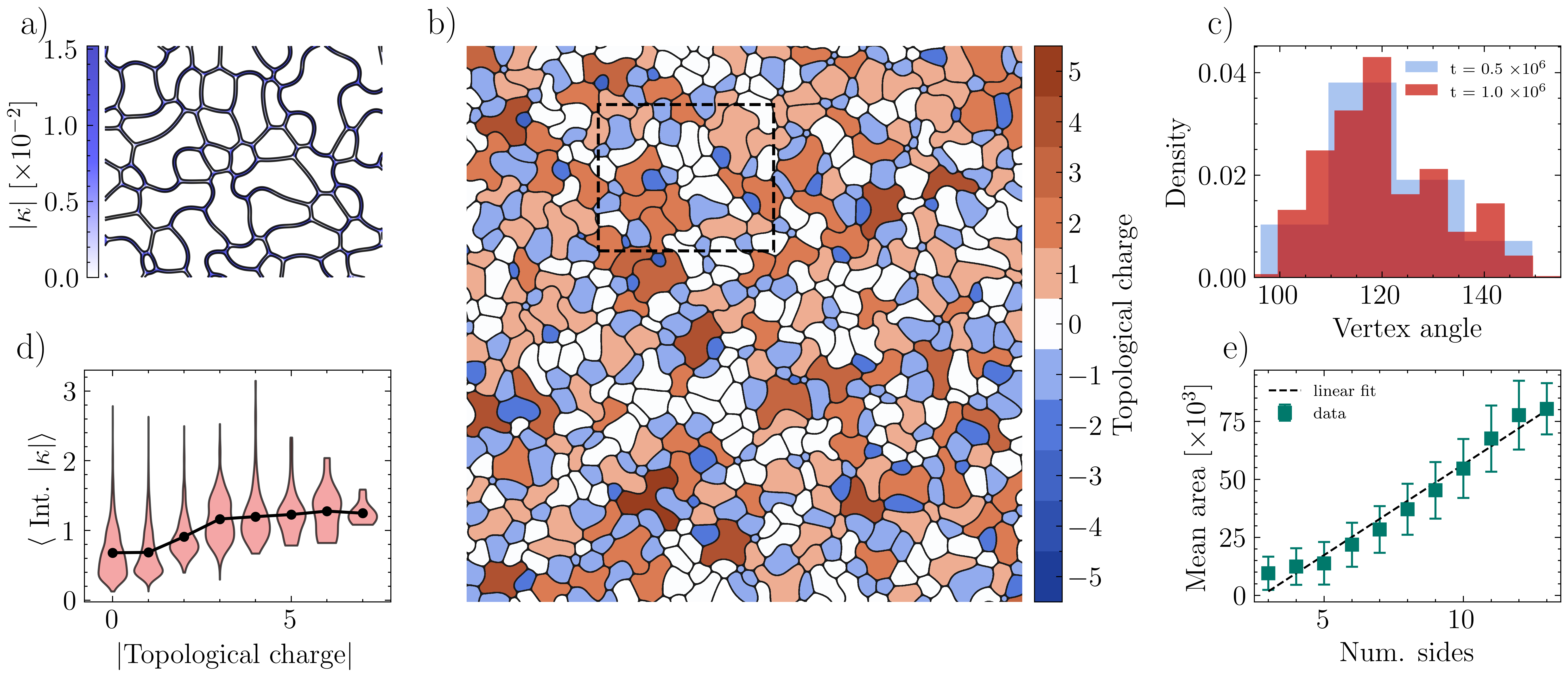}
    \caption{
    Non-equilibrium morphology of the network. (a) Magnified view of the network (the region marked by the dashed box in (b)) with edges coloured by absolute line curvature.
    (b) Large field-of-view snapshot with cells coloured by discrete topological charge $q = n - 6$, where $n$ is the number of bounding edges. (c) Distribution of vertex angles in degrees at three-fold junctions recorded at two well-separated times; the distribution is broad, and its mean is fixed at $120^\circ$ by the three-fold junction constraint. (d) Mean integrated absolute line curvature along a cell perimeter as a function of absolute topological charge $|q|$. (e) Mean cell area as a function of the number of sides. Simulations were performed until $t=10^6$, with data collected over three independent realisations.}
    \label{fig:topology}
    \end{center}
\end{figure*}

We begin by characterising the morphological states accessible to Active Model H$^{-}$ across the $(\lambda, \zeta)$ parameter space. Figure~\ref{fig:phase_diagram}(a) presents the resulting phase diagram. At low activity the system behaves qualitatively like the passive Cahn--Hilliard case, in which the minority phase forms coarsening droplets that coalesce and grow without bound via Ostwald ripening. Increasing both $\lambda$ and $\zeta$ drives a transition to a percolating filamentous network state, in which the characteristic domain width saturates (Fig.~\ref{fig:phase_diagram}(d)) while the network continues to rearrange slowly through local topological changes. The transition requires $\zeta\gtrsim5$ together with $\lambda\gtrsim5$--$10$, and at the resolution of our parameter grid the boundary is not monotonic in $\zeta$: networks appear already at $\lambda=5$ for $\zeta=20$, but only from $\lambda=10$ for $\zeta=5$ and $\zeta=50$. 
The transition into a network is captured by the isoperimetric ratio (Fig.~\ref{fig:phase_diagram}(b)), defined as $Q = \frac{4 \pi\,\mathrm{Area}}{\mathrm{Perimeter}^2}$%
with $Q=1$ for a perfect circle and $Q\rightarrow0$ indicating highly elongated structures.

The distinction between this network state and the foam-like structures produced by AMB$+$ is made explicit in Fig.~\ref{fig:phase_diagram}(c), which compares long-time snapshots of the passive Cahn--Hilliard equation, AMB$+$, and Active Model H$^{-}$ at $(\lambda,\zeta)=(10,5)$. While AMB$+$ produces structures that are connected in the phase field, magnified contours at $\phi_\mathrm{mid} = (\phi_+ + \phi_-)/2$ (inset) reveal that the AMB$+$ structure forms interstitial boundaries around closely packed droplets of the majority phase, rather than the genuine filamentous strands connected through nodes that characterise the network in our model.

For these filaments, Fig.~\ref{fig:phase_diagram}(d) shows that the mean filament width remains stable and approximately constant over long times, with the inset showing representative snapshots near the initial state, when filaments first become isolated, and at late times. Comparing the phase-field profiles between vertices, AMB$+$ shows a marked tapering away from the nodes, whereas Active Model H$^{-}$ retains a far flatter profile along the length of the strand (Fig.~\ref{fig:phase_diagram}(e)) and does not break for the parameter set presented here. This weaker depletion may allow filament edges in our model to persist over greater lengths, so that the resulting networks form stable out-of-equilibrium structures, which we investigate in the next section.

\subsection*{Non-equilibrium morphology of the network}

We now characterise the internal topology of the arrested network at $(\lambda,\zeta)=(10,50)$, and use these parameters for the rest of the paper. Figure~\ref{fig:topology}(b) shows a large field-of-view snapshot of the network at $t=10^6$, with individual cells coloured by their discrete topological charge $q = n - 6$, where $n$ is the number of edges bounding each cell. The charge distribution is spatially heterogeneous, with extended connected chains of cells of similar non-zero charge rather than a random distribution of isolated defects. This patchiness is a signature of long-range correlations.

\begin{figure*}[t!]
    \begin{center}
    \includegraphics[width=165mm]{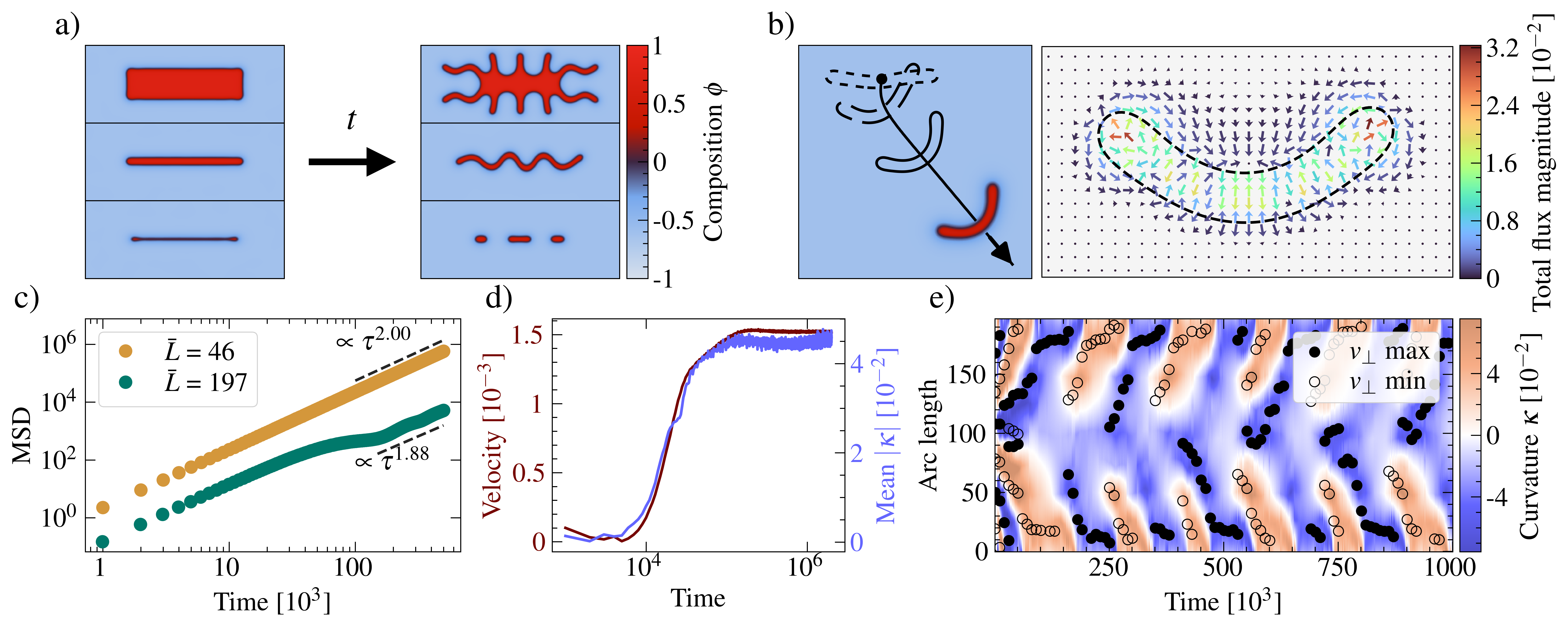}
    \caption{Curvature-driven self-propulsion of isolated filaments. (a) Snapshots of isolated filaments of widths $40$, $10$ and $4$ (top to bottom) subject to small perturbations, at an early time ($t=100$) and at later times ($t=(5,10,15)\times10^3$). (b) Left: snapshot of a short ($\overline{L}=46$) filament at $t=100\times10^3$, with outlines in dashed black indicating the time evolution ($t=(10,30,60)\times10^3$). Right: the coarse-grained total flux field $\boldsymbol{J}$ of the same filament at $t=2\times10^6$; arrows indicate the direction, colour the magnitude, and the dashed line identifies the contour at $\phi_\mathrm{mid}$. (c) Mean squared displacement of isolated filaments of two lengths on logarithmic axes, both showing superdiffusive exponents in the long-time limit. Simulation length $t=2\times10^6$. (d) Velocity magnitude of the centroid and mean absolute line curvature as functions of time for a short filament ($\overline{L}=46$). (e) Curvature kymograph of a long filament ($\overline{L}=197$), with the extrema of the perpendicular component of the filament velocity overlaid as black dots.}
    \label{fig:mechanism}
    \end{center}
\end{figure*}

We next examine individual edges in detail, using the line curvature $\kappa=\partial^2\boldsymbol{r}/\partial s^2  \cdot \hat{\mathbf{n}}$, with $\boldsymbol{r}$ denoting the position vector along the filament and $s$ its arc length. Figure~\ref{fig:topology}(a) reveals that the filaments forming the network edges are not the simple convex arcs that would be expected near equilibrium, but are frequently strongly curved and sometimes S-shaped. This anomalous edge geometry is quantitatively linked to the topological structure of the surrounding network: cells of larger absolute topological charge $|q|$ show systematically larger integrated absolute curvature along their perimeter (Fig.~\ref{fig:topology}(d)). The curvature of individual edges is therefore not independent of the network topology, but is coupled to it through the non-equilibrium dynamics.

At the junctions between edges, Fig.~\ref{fig:topology}(c) shows the distribution of vertex angles. Since the three angles meeting at a three-fold junction sum to $360^\circ$, the mean of this distribution is fixed at $120^\circ$ by construction and carries no information; the informative quantity is its width. The measured distribution is broad and asymmetric, spanning roughly $95^\circ$--$150^\circ$ with a pronounced shoulder near $130$--$140^\circ$, in contrast to the sharply peaked distribution expected for an interface-minimising equilibrium foam. Recording the distribution at two well-separated times shows essentially no change, so the junctions are stable and this spread does not decay over the accessible time window. We therefore attribute the angular disorder to the non-equilibrium dynamics rather than to an incomplete relaxation, while noting that our data establish stationarity over the accessible window rather than a strict separation from a very slowly relaxing passive foam.

Despite these locally non-equilibrium features, the network recovers a global statistical law characteristic of equilibrated cellular structures. Figure~\ref{fig:topology}(e) shows that the mean cell area scales linearly with the number of sides, in quantitative agreement with Lewis's law. The coexistence of persistent local disorder with macroscopic Lewis's law scaling reveals a competition between non-equilibrium driving at the scale of individual edges and emergent interface minimisation at the scale of the network as a whole.

\subsection*{Mechanism: curvature-driven self-propulsion of isolated filaments}
\begin{figure*}[t!]
\begin{center}
    \includegraphics[width=165mm]{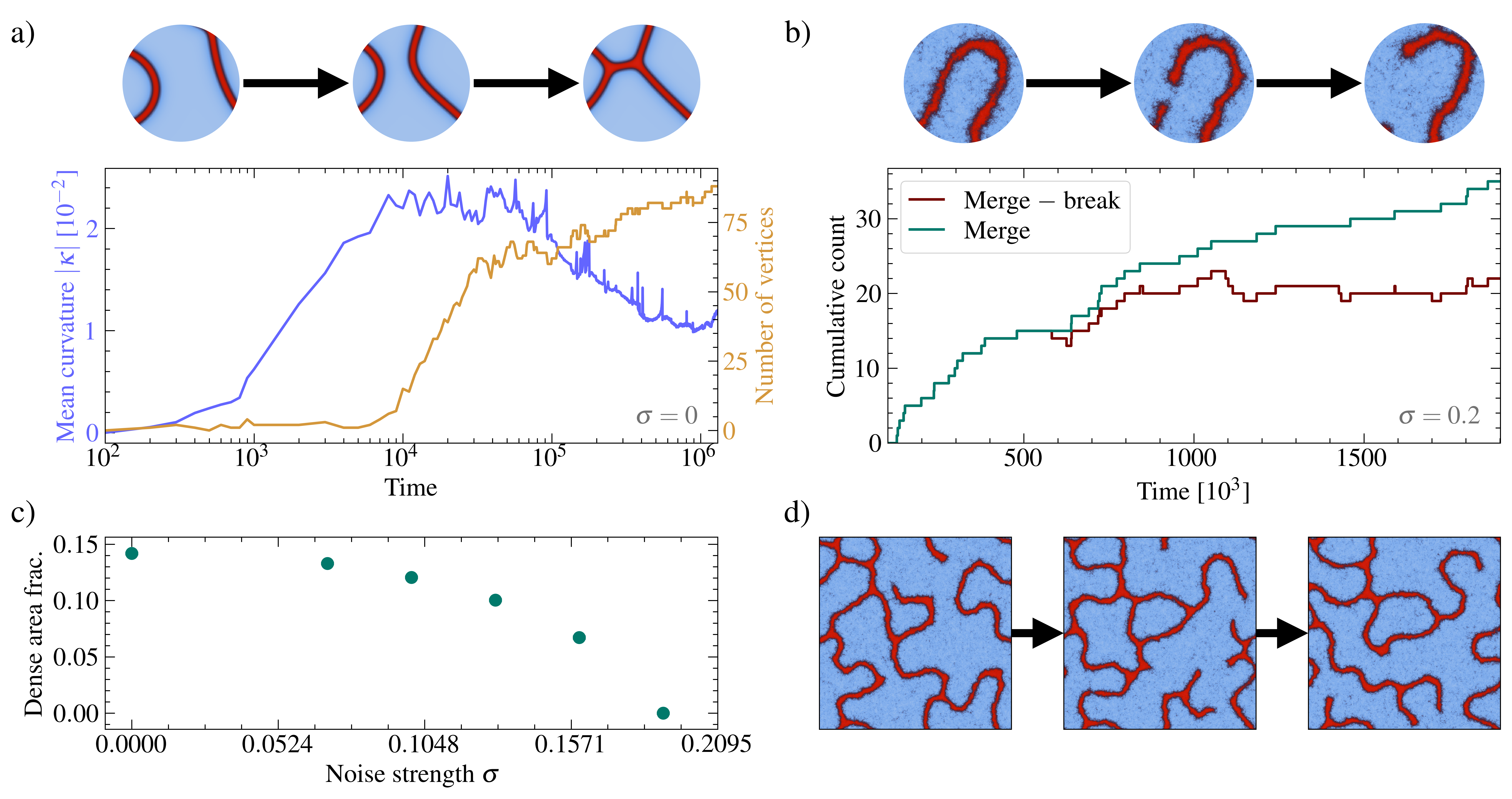}
    \caption{Curvature propulsion drives network assembly and break-and-reform dynamics under noise.
    (a) Top: Time-lapse of a representative merging event in which a filament develops a bend, propels toward a neighbouring filament, and forms a stable junction, shown for the deterministic case at elapsed times $t=(0, 225, 240)\times10^3$.
    Bottom: Mean curvature per filament (blue) and number of vertices (ochre) at early times, showing that curvature rises prior to branching and thus drives network formation.  (b) Top: Time-lapse of a representative break event for noise strength $\sigma=0.2$ at elapsed times $t=(0, 20, 170)\times10^3$.
    Bottom: Cumulative number of merge events (turquoise) and net merge events (merge events minus break events, red) show that for $\sigma=0.2$ break and merge events are at a steady state after an initial period. The simulation is started from a random field with $\phi_0 = -0.242$, which gives the same dense-phase area fraction as $\phi_0 = -0.4$ with the addition of noise.
    (c) The dense-phase area fraction decreases monotonically with $\sigma$. Simulations start from fully formed networks with initial concentration $\phi_0=-0.4$, where the phase boundary is taken from the deterministic case. (d) Snapshots of the simulation from (b) at $t= (1000, 1250, 1500)\times10^3$ illustrate dynamic network with break-and-reform mechanism.}
    \label{fig:assembly}
    \end{center}
\end{figure*}

The observations above raise an immediate question: what drives the assembly of isolated domains into a connected network, and what sustains the anomalous curvature of network edges over long times? To answer this we analyse the dynamics of individual isolated filaments, free from the constraints of the network.

We initialise straight filaments of varying width subject to small random perturbations and track their subsequent evolution. Figure~\ref{fig:mechanism}(a) summarises the width-dependent response. When initialised at smaller widths, filaments rupture into droplets or shorter segments, as the system attempts to conserve area while evolving towards a filament of preferred width. At larger initial widths, filaments undergo a spontaneous bending instability, deforming into curved shapes and ultimately developing capillary extensions from the interface. Similar capillary structures have been reported for AMB$+$~\cite{AMB+Capillary_Fausti2021}.

Alongside this instability, curved filaments develop an associated motility. Figure~\ref{fig:mechanism}(c) shows the mean squared displacement of isolated filaments of two different lengths. Short filaments display ballistic motion with a scaling exponent $\alpha \approx 2$, demonstrating directed self-propulsion rather than enhanced diffusion.
Longer filaments show a broader range of dynamics, including changes of direction, rotation and recombination into loops, and a correspondingly reduced exponent: the filament shown in Fig.~\ref{fig:mechanism}(c) enters a subdiffusive regime at intermediate times before recovering superdiffusive transport ($\alpha\approx1.88$) at long times.

This propulsion is curvature driven, as shown in Fig.~\ref{fig:mechanism}(d), which plots the temporal evolution of the short-filament velocity magnitude alongside its mean absolute line curvature. The onset of motility coincides closely with the development of curvature, indicating that propulsion is curvature driven rather than arising from another symmetry-breaking mechanism, such as active nematic driving~\cite{activefluids_formnetworks}.

This correspondence also holds for the longer filament, and is shown in the curvature kymograph of Fig.~\ref{fig:mechanism}(e),
where velocity extrema coincide with curvature extrema that nucleate near the filament centre and propagate outward to the free ends. As each extremum reaches a free end, it causes that end to pivot, which in turn promotes a new curvature extremum of opposite sign; this extremum then propagates outward in turn, reversing the direction of motion. This repeated back-and-forth pivoting gives rise to the subdiffusive regime observed at intermediate timescales. The duration of the subdiffusive regime could depend on the filament length relative to a characteristic wavelength: filaments shorter than this wavelength support only a single curvature extremum and therefore move in one direction only, recovering the ballistic behaviour of short filaments.

This link between curvature and motion arises from an asymmetric total flux $\boldsymbol{J}$ across the filament, shown in Fig.~\ref{fig:mechanism}(b). The flux is directed inward on the concave side and outward on the convex side, propelling the filament towards its convex side in a manner analogous to a sail driven by an asymmetric pressure field.

\subsection*{Curvature propulsion drives network assembly}

With the single-filament mechanism established, we return to the network and show directly that curvature-driven propulsion causes assembly. Figure~\ref{fig:assembly}(a) plots the mean curvature per filament and the number of vertices as functions of time. Where assembly proceeds through an intermediate filament stage, the mean curvature rises before the vertex count increases.
The corresponding time-lapse (Fig.~\ref{fig:assembly}(a), top) shows a straight filament developing a bend, propelling towards a neighbouring domain, and connecting to form a stable junction.

\subsection*{Balanced break and reform dynamics under noise}

Having investigated the deterministic assembly mechanism, we ask whether the resulting network state is robust to the introduction of additive Gaussian noise. Fluctuations act directly through the flux, driving continual small displacements of network edges, and indirectly by reducing the dense-phase area fraction, which decreases monotonically with noise strength $\sigma$ (Fig.~\ref{fig:assembly}(c)), consistent with a shift in the effective binodals. At sufficient noise strength the filaments rupture. Once a strand narrows enough, the curvature-induced pull separates the ends and prevents immediate reconnection, as shown in the time-lapse of Fig.~\ref{fig:assembly}(b, top) and the snapshots in Fig.~\ref{fig:assembly}(d). Rather than remaining fragmented, the free ends generated by rupture can develop curvature, propel towards neighbouring segments, and merge, restoring connectivity through the same curvature-propulsion mechanism responsible for the network's initial formation.

To test whether rupture and merging balance, we track the cumulative number of merge events and the net number of merge events (merges minus breaks) over time. Since the dense area fraction is reduced to 0 before breaking is observed, we start from a random initial field with $\phi_0=-0.242$ to match the dense area fraction of the deterministic case. The simulation shows that after an initial transient the net merge count plateaus (Fig.~\ref{fig:assembly}(b, bottom)), showing that break and merge events balance and proceed at a steady rate. The network at $\sigma=0.2$ is thus not static, and consists of continuous rupture and subsequent merge events, with the  rupture and reform rates balancing out.

The network state is therefore sustained by continuous non-equilibrium driving, is disrupted locally by fluctuations, and actively repairs itself through the same curvature-driven propulsion and merging that drive its initial assembly.

\section*{Discussion}

We have introduced Active Model H$^{-}$, a scalar phase-field theory of non-equilibrium phase separation in the overdamped hydrodynamic limit, and shown that it produces truly connected filamentous networks with arrested coarsening. The central result is a clean mechanistic picture: the active dynamics introduce a bending instability, the resulting curvature drives ballistic self-propulsion towards the convex side of the interface, and this motility causes isolated filaments to collide and assemble into a percolating network that shows break-and-reform dynamics once fluctuations are introduced. The network morphology is measurably non-equilibrium at the local scale yet recovers Lewis's law globally, pointing to a non-trivial competition between active driving and emergent interface minimisation.

The relationship between Active Model H$^{-}$ and existing scalar theories deserves careful discussion. AMB$+$ and the model introduced here are identical up to the $\phi$-dependence of the $\zeta$ term. They share the same spinodal structure and a similar coarsening phenomenology at low $\zeta$ and high $\lambda$. The key distinction is that the $\phi$-dependent advection couples asymmetrically to the interface curvature between the dense and dilute phases, and it is precisely this asymmetry that stabilises the connections between filaments by keeping the thickness of the edges uniform  and sharpening the vertices.

We propose that this operates as follows. In both cases the total flux draws dense material from the surrounding regions into the junction. In AMB$+$ the $\zeta$ term supports this motion, widening junctions at the expense of the adjacent edges, which become thinner and prone to rupture.
In Active Model H$^{-}$ the $\zeta$ term instead introduces an advective flux that counteracts this draw, keeping the connecting edges at a uniform thickness. This picture is consistent with the flat composition profile of Fig.~\ref{fig:phase_diagram}(e), but we have not measured the flux near a junction directly, and a systematic comparison of $\boldsymbol{J}$ around junctions in the two models would be needed to establish it. Whether this mechanism can be derived from a more microscopic picture of self-propelled particles near interfaces remains an interesting open question.

The network topology connects to a broader literature on cellular networks far from equilibrium. The coexistence of persistent vertex-angle disorder with global Lewis's law scaling has a direct precedent in multi-phase-field models of active cellular structures, which recover passive topological and geometrical scaling laws despite continuous non-equilibrium driving~\cite{wenzel2019topological}, and is reminiscent of reconstituted microtubule--kinesin ``active foams'', which form percolating cellular networks sustained by ongoing topological rearrangements~\cite{lemma2022active}. The present model offers a minimal setting in which this coexistence should be amenable to be studied analytically.

The observed break-and-reform dynamics (Fig.~\ref{fig:assembly}(b,d)) raise the question of whether sufficiently strong fluctuations can drive a loss of percolation, taking the system from a system-spanning network to a disconnected filament gas. An analogous activity-driven connectivity transition has been reported in reconstituted actomyosin gels, where motor activity reduces network connectivity towards a critical point~\cite{alvarado2013molecular}. A systematic study of this transition as a function of $\sigma$, $\lambda$ and $\zeta$ would establish whether it is continuous or discontinuous, and whether it belongs to a known universality class. The reduction in dense-phase area fraction with noise strength (Fig.~\ref{fig:assembly}(c)) hints at a renormalisation of the effective phase boundary by fluctuations, which could be investigated using the tools developed for stochastic active field theories~\cite{review_Nardini2022}.

Several extensions of the model are natural next steps. The present study is restricted to two dimensions; extending it to three dimensions would allow comparison with those observed in experimental soft matter, and would likely introduce new topological features associated with the higher dimensionality. Additionally, coupling Active Model H$^{-}$ to a momentum-conserving fluid and recovering the full Active Model H would allow the role of hydrodynamic interactions in network assembly to be assessed~\cite{padhan_suppression_2025}. Since the curvature-driven flux considered here is local and interfacial, we expect momentum conservation to matter chiefly by introducing long-ranged interactions between filaments, and hence a hydrodynamic screening length that has no counterpart in the overdamped limit.

More broadly, the results presented here suggest that truly connected filamentous networks are accessible to purely scalar active field theories. This is a weaker requirement than the introduction of a tensorial order parameter, and it opens the possibility that scalar models may be sufficient to describe a wider class of non-equilibrium
network-forming systems than previously thought.



\subsection*{Acknowledgements} 

A. D. acknowledges funding from the Novo Nordisk Foundation (grant No. NNF18SA0035142 and NERD grant No. NNF21OC0068687), Villum Fonden (Grant No. 29476), and the European Union (ERC, PhysCoMeT, 101041418). Views and opinions expressed are however those of the authors only and do not necessarily reflect those of the European Union or the European Research Council. Neither the European Union nor the granting authority can be held responsible for them. The Tycho supercomputer hosted at the SCIENCE HPC center at the University of Copenhagen was used to support this work.

\bibliographystyle{ieeetr}
\bibliography{references}
\end{document}